\documentclass[%
 reprint,
groupedaddress,
showpacs,
 amsmath,amssymb,
 aip,
]{revtex4-1}

\usepackage[final]{graphicx}
\usepackage{dcolumn}
\usepackage{bm}

\begin{document}


\title{Stability and electronic structure of covalently functionalized graphene layers}

\author{Karolina Z. Milowska}
 \email{karolina.milowska@gmail.com}
\author{Jacek A. Majewski}%
 \email{Jacek.Majewski@fuw.edu.pl }
\affiliation{%
 Institute of Theoretical Physics, Faculty of Physics, University of Warsaw, ul. Ho\.za 69,
PL-00-681 Warszawa, Poland
\\
}%

\date{\today}

\begin{abstract}
We present exemplary results of extensive studies of mechanical, electronic and transport properties of covalent functionalization of graphene monolayers (GML) with -NH$_2$. We report new results of \textit{ab initio} studies of covalent functionalization of GML with -NH$_2$  groups up to 12.5$\%$ concentration. Our studies are performed in the framework of the density functional theory (DFT) and non-equilibrium Green's function (NEGF). We discuss the stability (adsorption energy), elastic moduli, electronic structure, band gaps, and effective electron masses as a function of the density of the adsorbed molecules. We also show the conductance and I(V) characteristic of these systems.
 Generally, the stability of the functionalized graphene layers decreases with the growing concentration of attachments and we determine the critical density of the molecules that can be chemisorbed on the surface of GLs. Because of local deformations of GLs and sp$^3$ rehybridization of the bonds induced by fragments, elastic moduli decrease with increasing number of groups. Simultaneously, we observe that the functionalizing molecules stretch the graphene's lattice, the effect being more pronounced for higher concentration of adsorbed molecules. We find out that the GLs functionalization leads in many cases to the opening of the graphene band gap (up to 0.5302 eV for 12.5$\%$ concentration) and can be therefore utilized in graphene devices. The new HOMO and LUMO originate mostly from the impurity bands induced by the functionalization and they exhibit parabolic dispersion with electron effective masses comparable to ones in silicon or gallium nitride. 
\end{abstract}


\maketitle   

\section{Introduction}

Nowadays graphene, a 2-dimensional semimetalic monolayer, a form of sp$^2$ hybridizated carbon, attracts a lot of research activity owing to its  unique electronic, mechanical and thermal properties \cite{a1}. Graphene layers (GLs) are emerging as the very promising candidates for a new generation of electronic devices \cite{a16,a17,a18,a26}. However, the GML has zero energy band gap. This hinders direct application of graphene layer in field effect transistors (FETs) and the functionalization of GLs could be a remedy for this problem.  The possibility to generate controllable band gap in graphene without significant deterioration of the remaining advantageous properties is desirable. Furthermore, covalent functionalization enables usage of graphene monolayers as sensors \cite{a26,a19}.

We have performed extensive studies of the stability, electronic  and transport properties of the functionalized graphene with various molecular groups \cite{m1,m2}. In this article we focus on the problem how the properties of GMLs change after the functionalization with -NH$_2$.

\section{Calculation details}

Our studies of GML functionalized with -NH$_2$ are based on {\sl ab initio} calculations within the framework of the spin polarized density functional theory (DFT) \cite{a3,a4} as implemented in the SIESTA package \cite{a6,a7}. The so-called PBE form of the generalized gradient approximation (GGA) \cite{a5} is chosen for the exchange correlation density functional. The computations involving full geometry optimization have been performed employing following parameters of the SIESTA package that determine numerical accuracy of the results: double-$\zeta$-plus-polarization basis, kinetic energy mesh cutoff of 350 Ry, the self-consistency mixing rate of 0.1, the convergence criterion for the density matrix of 10$^{-4}$, maximum force tolerance equal to 0.01 eV/ \AA. Calculations were performed for supercell geometry with graphene layers separated by a distance large enough to eliminate all kinds of spurious interactions. For the band structure and density of states calculations, the Mesh Cutoff has been increased up to 550 Ry. For better imaging of density of states, the peak width for broadening the energy eigenvalues has been set to 0.05 eV. To investigate the various densities of the functionalizing amines, we have taken large lateral unit cells that contain 4, 9, 16, 25, and 36 standard two-atomic graphene unit cells arranged respectively in the patterns (2x2), (3x3), (4x4), (5x5), (6x6) (i.e., these supercells contain 8, 18, 32, 50 and 72 carbon atoms, respectively) and placed a functionalizing group in the enlarged unit cells.

The adsorption energy, which has been calculated according to the formula:
\begin{equation}
E_{ads}= (E_{tot}(GML+NH_{2}) - E_{tot}(GML)- E_{tot}(NH_{2}))
\end{equation}
is used as the measure of the stability of the systems studied. $E_{tot}(GML+NH_{2})$ is the total energy of functionalized graphene, $E_{tot}(GML)$ is the total energy of graphene monolayer, and $E_{tot}(NH_{2})$ is the total energy of the amine group.

Elastic properties such as Poisson ratio, Young's, shear and Bulk moduli have been also calculated.
The Poisson ratio values, has been obtained according to the equation:	
\begin{equation}
\label{e4}
\nu  =  - \frac{{\Delta a}}{a}\frac{b'}{{\Delta b'}},
\end{equation}	
where $\Delta a(b')$	is change of lattice constant(projection of lattice constant $b$ on perpendicular direction to lattice constant $a$) in strained GML. 
The most interesting quantity, Young's modulus, have been determined - from components ($\sigma_{ii}$) of the stress tensor:
\begin{equation}
\label{e22}
Y = \frac{{\sigma _{ii} }}{{\varepsilon _{ii} }},
\end{equation}
where $\varepsilon _{ii}$ is strain.
Bulk modulus (Eq.~\ref{e5}) and Shear modulus (Eq.~\ref{e6}), which could be easily derived from (Eq.~\ref{e4}) and (Eq.~\ref{e22}) are given by the following mathematical formula:
\begin{equation}
\label{e5}
B  =  \frac{Y}{3(1-2 \nu)},
\end{equation}
\begin{equation}
\label{e6}
S  =  \frac{Y}{2(1+ \nu)},
\end{equation}

We have used TranSIESTA \cite{a20} to study electronic coherent transport in the systems employing NEGF technique, within the Keldysh formalism, a rather standard procedure for treatment of coherent transport \cite{a21}. The structures have been treated as two-probe systems with the central scattering region sandwiched between semi-infinite source (left) and drain (right) electrode regions (as depicted in Fig.~\ref{transport}(a)).
The conductance can be expressed as follows \cite{a21}:
\begin{equation}
\label{eq2}
G(E)=G_o Tr[\Gamma_R G^r \Gamma_L G^a],
\end{equation}
where $G_o = 2\frac{e^2}{h}$ is the unit of quantum conductance and G$^{r(a)}$  is retarded (advanced) Green's function. The current through the scattering region has been calculated according to Landauer-Buttiker formula, i.e., assuming the limit of small bias, which is justified for the range of the external voltage considered in the present study: 
\begin{equation}
\label{eq3}
I(V_{DS}) = \int\limits_{\mu_R}^{\mu_L} {T(E,V_{DS} )dE},
\end{equation}
where V$_{DS}$ is equal to the electrochemical potential difference between the left and right electrodes (eV$_{DS}$=$\mu_L - \mu_R $) and T is the transmission.

\section{Results and discussion}

\begin{figure}[t]%
\includegraphics*[width=\linewidth]{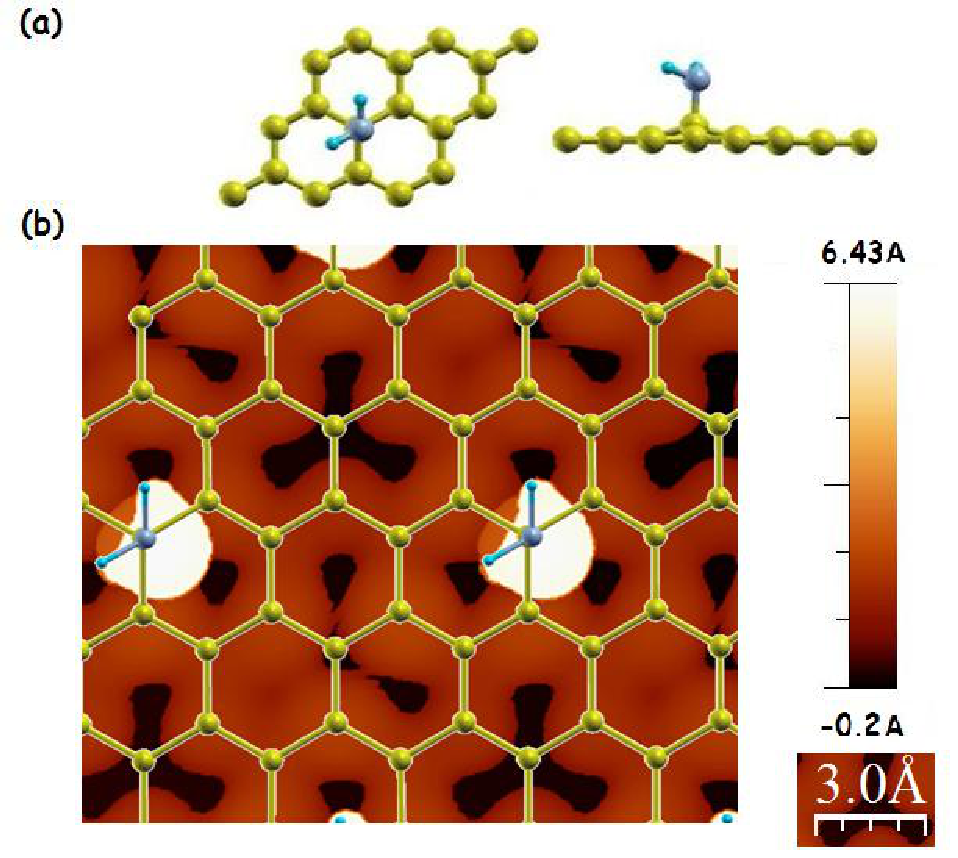}
\caption{%
(a) The optimized graphene monolayer functionalized with one -NH$_2$ group per (3x3) supercell. (b) Simulated STM image of structure visualised  above (V$_{bias}$=1V). Also superposed is a ball-and-stick model of the functionalized graphene lattice.}
\label{struk}
\end{figure}

Amine groups bind covalently to the graphene layer, implying sp$^2$ to sp$^3$ rehybridization and local deformation of the graphene plane, which are well known effects observed in chemical doping graphene layers and nanoribbons\cite{a26,a25,a22,a23}. C-C bond length in the closest proximity of the functionalizing group increases to 1.51 \AA, whereas C-C bond lengths away from the group are very close to the bond length in pure graphene (1.43 \AA). The carbon atom that is directly bonded to the group sticks out from the graphene sheet (see Fig.~\ref{struk} (a)), the effect being more pronounced for higher concentration of adsorbed molecules. 
The STM image depicted in Fig.~\ref{struk} (b) is calculated under the Tersoff-Hamann theory \cite{b2} and visualized using WSxM \cite{b1}. It shows bright spots corresponding to amine groups above graphene lattice. It is clearly seen that amines break the hexagonnal symmetry of graphene layer and create scattering centers. This is consistent with recent experimental work \cite{a25}.

The adsorption energy, used as a measure of the stability, decreases with growing concentration of groups. Because of local deformations of GLs and sp$^3$ rehybridization of the bonds induced by fragments, elastic moduli diminish with increasing number of groups. The highest reduction is equal to 15.5$\%$ in Shear, 14$\%$ in Young's and 9.5$\%$ in Bulk modulus (see Tab.~\ref{tabela}). In comparison to standard materials, the elastic properties of amine functionalized GMLs are still superior and its potential usage in solar cell is secured \cite{a26}.

\begin{figure}[t]%
\includegraphics*[width=\linewidth]{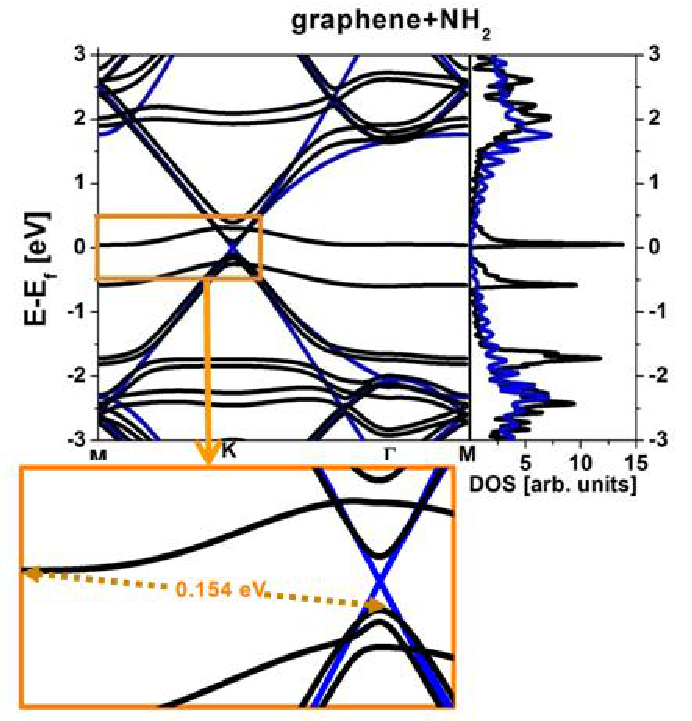}
\caption{%
Band structure and DOS of graphene functionalized with amine molecule (one per (3x3) supercell) are plotted with black lines, whereas these quantities for pristine graphene are plotted with blue dotted lines. Inset: zoom of band structure around Fermi level (shifted to 0) along direction M$\to$K in the Brillouin Zone. Indirect band gap is marked by the yellow dotted arrow.}
\label{bandDOS}
\end{figure}
We find out that the GLs functionalization leads in many cases to the opening of the graphene band gap (up to 0.5302 eV for 12.5$\%$ concentration - see Tab.~\ref{tabela}) and can be utilized in graphene devices. \footnote{The fundamental band gap has been calculated as the difference between the Kohn–-Sham energy of the highest occupied molecular orbital (HOMO) and the Kohn–-Sham energy of the lowest unoccupied molecular orbital (LUMO).} Similar results, but for bilayer graphene, were obtained by Boukhvalov \cite{a22}.

 The fundamental band gaps for the functionalized structures generally inrease with increasing concentration of amines and typically are indirect. It is specified in Fig.~\ref{bandDOS}.  The new HOMO and LUMO originate from the impurity bands induced by the functionalization and exhibit parabolic dispersion. Those impurity levels result mostly from nitrogen atom and $\pi$-bonds. The effective electron masses (m$_{eff}$) and mobility depend rather strongly on the concentration of dopants. 

\begin{table*}[htb]
  \caption{Graphene monolayers functionalized with amines: c- concentration of groups, E$_{ads}$ - adsorption energy, Y- Young's modulus, S - Shear modulus, B- Bulk modulus, $\eta$ - Poisson ratio, E$_{gap}$ -fundamental band gap, m$_{eff} ^{K-M}$ and m$_{eff} ^{K-\Gamma}$ - effective mass of electron in K point along the direction K-M and K-$\Gamma$, respectively.}
  \centering
  \begin{tabular}{@{}llllllllll@{}}
    \hline
     supercell & c [$\%$]  & E$_{ads}$ [eV] & Y [TPa] & S [TPa] & B [TPa] & $\eta$ & E$_{gap}$ & m$_{eff} ^{K-M}$ [m$_e$] & m$_{eff} ^{K-\Gamma}$ [m$_e$] \\
    \hline
    (6x6)  & 1.39  & -0.883 & 1.02	 & 0.41	& 0.65 & 	0.24 &	0.0058 &	0.005 &	0.010  \\
   (5x5) &	2.00 &	-0.846 &	1.03 &	0.43 &	0.57 &	0.20 &	0.0732 &	2.675	& 2.774\\
   (4x4) &	4.35&	-0.825 &	1.01 &	0.42 &	0.54 &	0.19 &	0.0852 &	3.368 &	6.144 \\
(3x3)	& 5.56 &	-0.799 &	0.98 &	0.41 &	0.56 &	0.20 &	0.1535 &	0.038	 & 0.039 \\
(2x2) &	12.5 &	-0.739 &	0.90 &	0.38 &	0.46 &	0.17 &	0.5302 &	20.148 &	20.720 \\
    \hline
  \end{tabular}
  \label{tabela}
\end{table*}

\begin{figure}[t]%
\includegraphics*[width=\linewidth]{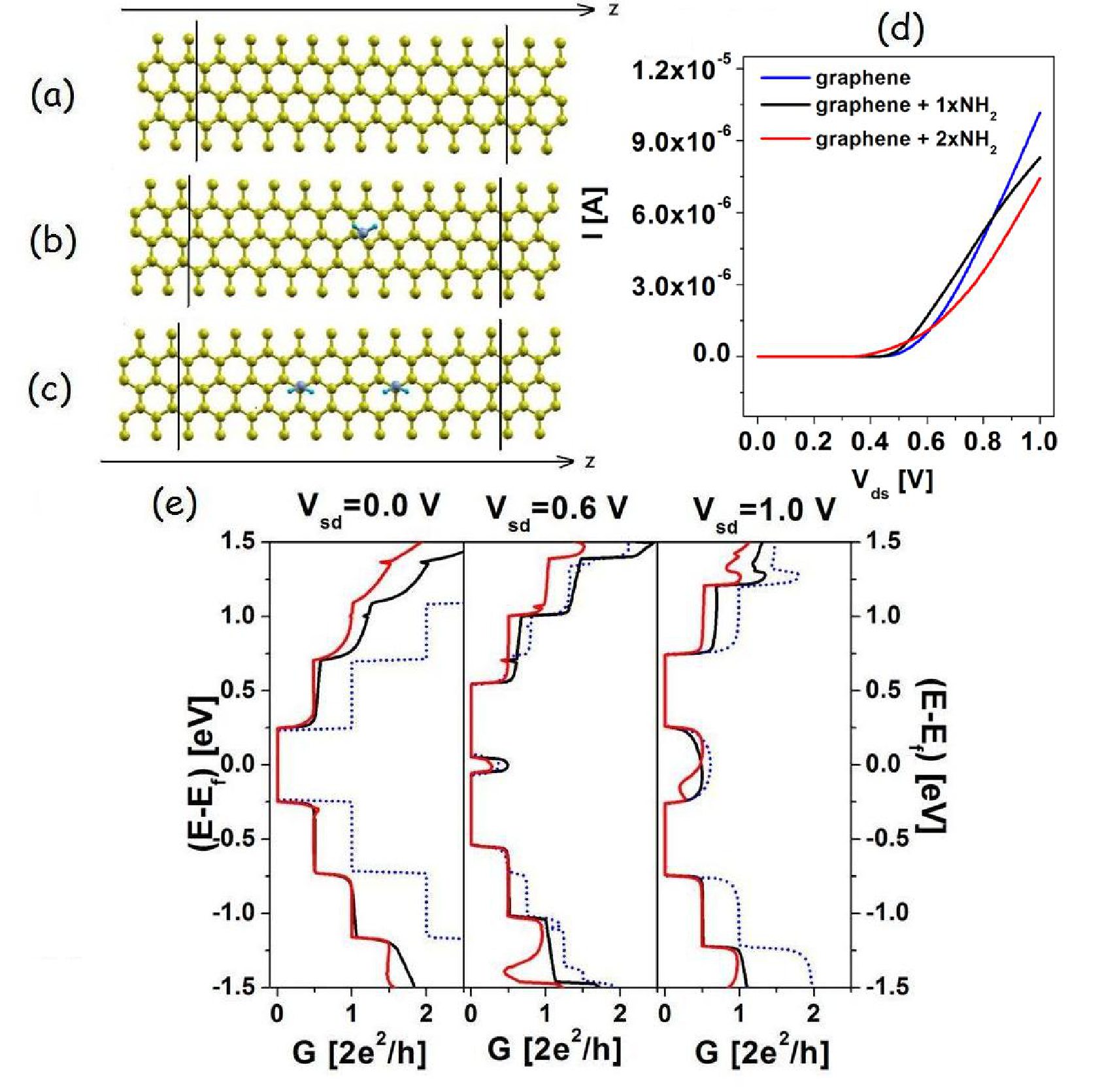}
\caption{%
Three systems studied for transport properties: pure graphene monolayer (a), graphene functionalized with one (b) and two (c) NH$_2$ groups. Electrode regions contain 16 carbon atoms, whereas scattering regions contains 80 carbon atoms and zero, one, and two -NH$_2$ groups for cases (a), (b), and (c), respectively. Periodic boundary conditions are conserved in perpendicular direction to the z axis. The voltage is applied along the z axis. (d) I(V) characteristic for pure (blue dotted line) and functionalized graphene with one (black solid line) and two (red solid line) -NH$_2$ groups. (e) The transmission spectra for three different source-drain voltages calculated for three systems described above.}
\label{transport}
\end{figure}

At some concentrations, they are comparable to masses in silicon or gallium nitride (see Tab.~\ref{tabela}). 
For (3x3) and (6x6) supercells m$_{eff}$ are close to 0, which is consistent with the phenomenon of zero band gap for supercells with dimensions being multiples of three graphene lattice constants described by Garcia-Lastra\cite{a10}. Breaking of the hexagonal symmetry of the layer is also clearly seen in effective masses. They are typically slightly bigger along K-$\Gamma$ than K-M direction. 
   
From the transport calculations (see Fig.~\ref{transport}), one can see that this type of functionalization hardly changes the conductance (Fig.~\ref{transport}(e)) and I(V) characteristic (Fig.~\ref{transport}(d)) in comparison to pure graphene monolayer.
Nevertheless, it is clearly seen that the conduction around Fermi level decreases with increase of concentration of adsorbants. The critical voltage for the non-zero current is the lowest for the system with two amines and the highest for the pristine graphene. Also for low bias, the current for functionalized systems is larger than for unfunctionalized one. However, for the bias larger  than 0.6 V, we observe the largest current for pristine graphene and the smallest for the system with higher concentration of functionalizing amines. Let us remark that the conductance of graphene functionalized with amines drops much slowler with concentartion of dopants than in the case of functionalization with hydroxyl or carboxyl groups. It is mostly due to the fact that nitrogen acts as donor. 
This is consistent with the experimental results of Baraket \cite{a25}. The increasing concentration of primary amines covalently bound to graphene monolayer causes increase in the chemical reactivity of the surface, while the electrical conductivity decreased.  However, even highly aminated graphene, up to 20$\%$, is conductive enough to be be used for DNA detection as bio-attachment platform in a biologically active field-effect transistors. 
Therefore, it is plausible that these functionalized structures can be commercially used as biosensors \cite{a19,a26}.

As it was pointed out by Garcia-Lastra\cite{a27}, the behavior of the transmission $G$ as depicted in Fig.~\ref{transport} is characteristic for metallic carbon nanotubes with single adsorbed molecule placed on top of carbon atom, or more adsorbed molecules placed on equivalent lattice sites (say $A$) with certain positions determined by the vectors $\vec{R}=n \cdot \vec{a} + m \cdot \vec{b}$, with $ n - m = 3p$, where $m$, $n$, and $p$ are integers, and $\vec{a}$ and $\vec{b}$ are two primitive translations of the graphene lattice. In the case studied in the present paper, the two -NH$_2$ molecules were placed at graphene lattice sites with $(n,m)$ equal to $(0,0)$ and $(3,0)$ at so-called top positions, therefore, fulfilling the rule\cite{a27} and indicating that graphene layer can be treated as metallic carbon nanotube of infinite radius. We have undertaken the further studies of the coherent transport in covalently functionalized GML to find out some further possible analogies to the rules presented in Ref.\onlinecite{a27}, which will be published elsewhere.

\section{Conclusions}

In summary, we have performed {\sl ab initio} calculations of the mechanical, electronic and transport  properties of the graphene monolayer functionalized with -NH$_2$ molecules. Covalent functionalization changes the local and global structure of the functionalized system, causes sp$^2$ $\to$ sp$^3$ rehybridization, reduces the elastic moduli, and introduces impurity levels around Fermi energy, which allows us for band gap engineering. This type of functionalization slightly reduces current and only moderately modifies the I(V) characteristics in comparison to the pristine graphene.

\section{Acknowledgement}
This work has been supported by the European Founds for Regional Development within the SICMAT Project (Contact No. UDA-POIG.01.03.01-14-155/09). We acknowledge also support of the PL-Grid Infrastructure and of Interdisciplinary Centre for Mathematical and Computational Modeling, University of Warsaw (Grant No. G47-5).


\begin{thebibliography}{[1]}

\bibitem{a1}
A.\,K. Geim, K.\,S. Novoselov,
Nat. Mater.  \textbf{6}, 183 (2007).

\bibitem{a16}
S.~Park, R.\,S. Ruoff,
Nat. Nanotechnol. \textbf{4}, 217 (2009).

\bibitem{a17}
D.~Pacile, J.\,C. Meyer, A.~Fraile Rodryguez, M.~Papagno, C.~Gomez-Navarro, R.\,S. Sundaram, M.~Burghard, K.~Kern, C.~Carbone, U.~Kaiser,
Carbon \textbf{49}, 966 (2011).

\bibitem{a18}
R.~Sengupta, M.~Bhattacharyaa, S.~Bandyopadhyayb, A.\,K. Bhowmick,
Prog. Polym. Sci. \textbf{36}, 638 (2011).

\bibitem{a26}
V.~Georgakilas~, M.~Otyepka, A.\,B. Bourlinos, V.~Chandra, N.~Kim, K.\,Ch. Kemp, P.~Hobza, R.~Zboril, K.\,S. Kim,
Chem. Rev. \textbf{112}, 6156 (2012).

\bibitem{a19}
Y.~Shao, J.~Wang, H.~Wu, J.~Liu, I.\,A. Aksay, Y.~Lina,
Electroanalysis \textbf{22}, 1027 (2010).

\bibitem{m1}
K.~Milowska, M.~Birowska, J.\,A. Majewski,
Acta Physica Polonica A \textbf{120}, 842 (2011).

\bibitem{m2}
K.~Milowska, M.~Birowska, J.\,A. Majewski,
Diamond and Related Materials \textbf{23}, 167 (2012). 

\bibitem{a3}
P.~Hohenberg, W.~Kohn, 
Phys. Rev. \textbf{136}, 864 (1964).

\bibitem{a4}
W.~Kohn, L.\,J. Sham, 
Phys. Rev. \textbf{140}, A1133 (1965).


\bibitem{a6}
D.~Sanchez-Portal, P.~Ordejon, E.~Artacho, J.\,M. Soler, 
Int. J. Quantum Chem. \textbf{65}, 453 (1997).

\bibitem{a7}
J.\,M. Soler, E.~Artacho, J.~Gale, A.~Garcia, J.~Junquera, P.~Ordejon, D.~Sanchez-Portal, 
J. Phys.:Condens. Matter \textbf{14}, 2745 (2002).

\bibitem{a5}
J.\,P. Perdew, K.~Burke, M.~Ernzerhof, 
Phys. Rev. Lett. \textbf{77}, 3865 (1996).

\bibitem{a20}
M.~Brandbyge,  J.\,L.  Mozos, P.~Ordejon, J.~Taylor, K.~Stokbro, 
Phys. Rev. B \textbf{65}, 165401 (2002).


\bibitem{a21}
S.~Datta, 
Electronic Transport in Mesoscopic Systems, Cambridge University Press, 1995.

\bibitem{a25}
M.~Baraket, R.~Stine, W.\,K. Lee, J.,T. Robinson, C.\,R. Tamanaha, P.\,E. Sheehan, S.\,G. Walton,
Appl. Phys. Lett. \textbf{100}, 233123 (2012).


\bibitem{a22}
D.\,W. Boukhvalov., M.\,I. Katsnelson,
Phys. Rev. B \textbf{78}, 085413 (2008).

\bibitem{a23}
D.\,W. Boukhvalov., M.\,I. Katsnelson,
J. Phys.: Condens. Matter \textbf{21},344205 (2009). 

\bibitem{b2}
J. Tersoff and D.R. Hamman,
Phys. Rev. Lett. \textbf{50}, 1998 (1983).

\bibitem{b1}
I.~Horcas, R.~Fernandez, J.-M. Gomez-Rodrigez, J. Colchero, J. Gomez-Herrero, A.\,M. Baro,
Rev. Sci. Instrum. \textbf{78}, 013705 (2011).

\bibitem{a10}
J.\,M. Garcia-Lastra,
Phys. Rev. B \textbf{82}, 235418 (2010).

\bibitem{a27}
J.\,M. Garcia-Lastra, K.\,S. Thygesen, M. Strange, and A. Rubio,
Phys. Rev. Lett. 101, 236806 (2008).

\end{thebibliography}
\end{document}